\documentclass[aps,prc,twocolumn,superscriptaddress,floatfix,amsmath,amssymb,nofootinbib,longbibliography]{revtex4-1}
\usepackage[utf8]{inputenc}
\usepackage[T1]{fontenc}

\usepackage{xcolor}
\usepackage{graphicx}
\usepackage[citecolor=blue,linkcolor=blue,urlcolor=blue]{hyperref}
\usepackage[ocgcolorlinks]{ocgx2}
\usepackage{scalefnt}
\usepackage{xspace}
\usepackage{times,txfonts}
\usepackage{pifont}
\usepackage{bm}

\newcommand{\la}{\langle}
\newcommand{\ra}{\rangle}

\newcommand{\ai}{\emph{ab initio}}

\newcommand{\Xmax}[1]{#1_\text{max}}
\newcommand{\Xmin}[1]{#1_\text{min}}
\newcommand{\kappaX}[1]{\kappa^{#1}}

\begin{document}

\title{Importance truncation for the in-medium similarity renormalization group}

\author{J.~Hoppe}
\email{jhoppe@theorie.ikp.physik.tu-darmstadt.de}
\affiliation{Technische Universit\"at Darmstadt, Department of Physics, 64289 Darmstadt, Germany}
\affiliation{ExtreMe Matter Institute EMMI, GSI Helmholtzzentrum f\"ur Schwerionenforschung GmbH, 64291 Darmstadt, Germany}

\author{A.~Tichai}
\email{alexander.tichai@physik.tu-darmstadt.de}
\affiliation{Technische Universit\"at Darmstadt, Department of Physics, 64289 Darmstadt, Germany}
\affiliation{ExtreMe Matter Institute EMMI, GSI Helmholtzzentrum f\"ur Schwerionenforschung GmbH, 64291 Darmstadt, Germany}
\affiliation{Max-Planck-Institut f\"ur Kernphysik, Saupfercheckweg 1, 69117 Heidelberg, Germany}

\author{M.~Heinz}
\email{mheinz@theorie.ikp.physik.tu-darmstadt.de}
\affiliation{Technische Universit\"at Darmstadt, Department of Physics, 64289 Darmstadt, Germany}
\affiliation{ExtreMe Matter Institute EMMI, GSI Helmholtzzentrum f\"ur Schwerionenforschung GmbH, 64291 Darmstadt, Germany}
\affiliation{Max-Planck-Institut f\"ur Kernphysik, Saupfercheckweg 1, 69117 Heidelberg, Germany}

\author{K.~Hebeler}
\email{kai.hebeler@physik.tu-darmstadt.de}
\affiliation{Technische Universit\"at Darmstadt, Department of Physics, 64289 Darmstadt, Germany}
\affiliation{ExtreMe Matter Institute EMMI, GSI Helmholtzzentrum f\"ur Schwerionenforschung GmbH, 64291 Darmstadt, Germany}
\affiliation{Max-Planck-Institut f\"ur Kernphysik, Saupfercheckweg 1, 69117 Heidelberg, Germany}

\author{A.~Schwenk}
\email{schwenk@physik.tu-darmstadt.de}
\affiliation{Technische Universit\"at Darmstadt, Department of Physics, 64289 Darmstadt, Germany}
\affiliation{ExtreMe Matter Institute EMMI, GSI Helmholtzzentrum f\"ur Schwerionenforschung GmbH, 64291 Darmstadt, Germany}
\affiliation{Max-Planck-Institut f\"ur Kernphysik, Saupfercheckweg 1, 69117 Heidelberg, Germany}

\begin{abstract}
\textit{Ab initio} nuclear many-body frameworks require extensive computational resources, especially when targeting heavier nuclei. Importance-truncation (IT) techniques allow one to significantly reduce the dimensionality of the problem by neglecting unimportant contributions to the solution of the many-body problem. In this work, we apply IT methods to the nonperturbative in-medium similarity renormalization group (IMSRG) approach and investigate the induced errors for ground-state energies in different mass regimes based on different nuclear Hamiltonians. We study various importance measures, which define the IT selection, and identify two measures that perform best, resulting in only small errors to the full IMSRG(2) calculations even for sizable compression ratios. The neglected contributions are accounted for in a perturbative way and serve as an estimate of the IT-induced error. Overall we find that the IT-IMSRG(2) performs well across all systems considered, while the largest compression ratios for a given error can be achieved when using soft Hamiltonians and for large single-particle bases.
\end{abstract}

\maketitle

\allowdisplaybreaks

\section{Introduction}

The description of nuclear many-body systems using \ai{} methods has undergone substantial progress in recent years, extending the reach to heavier and more exotic systems as well as to electroweak observables~\cite{Hergert2020review,Stro21atomicNucl,Gysb19beta,Kauf2068Ni,Friman-Gayer:2020vqn,Nova21CC0v2b}, and investigating sophisticated uncertainty quantification methods~\cite{Furn15uncert,LENPIC:2018lzt,Phil21BAND}.
In the past, the nuclear chart was mostly accessible from phenomenological approaches like the nuclear shell model or energy density functional theory, whereas \ai{} approaches were limited to light systems up to mass number $A\lesssim 16$ using large-scale diagonalization configuration interaction (CI) (see, e.g., Ref.~\cite{Barr13PPNP}) approaches or quantum Monte Carlo methods~\cite{Carl15RMP,Lonardoni:2017hgs}.
Due to their intrinsic computational scaling, the extension to heavier systems was computationally not feasible within the above frameworks.
This picture changed drastically with the proliferation of many-body expansion techniques that employ a suitably chosen $A$-body reference state as starting point, e.g., a Slater determinant obtained from a Hartree-Fock (HF) calculation.
Residual correlation effects on top of the reference state are accounted for through a correlation expansion.
Various frameworks exist for the design of such expansion schemes, e.g., many-body perturbation theory (MBPT)~\cite{Holt14Ca,Tich16HFMBPT,Tichai18BMBPT,Tichai2020review}, coupled-cluster (CC) theory~\cite{Hage14RPP,Bind14CCheavy}, self-consistent Green's function (SCGF) theory~\cite{Dick04PPNP,Soma20SCGF}, and the in-medium similarity renormalization group (IMSRG)~\cite{Tsuk11IMSRG,Herg16PR,Stroberg2019}.
Once truncated to a given order in the many-body expansion, these methods allow mild polynomial scaling, in contrast to the exponential scaling of exact methods, thus providing access to heavier systems.
This has led to the \ai{} calculation of systems with up to approximately one hundred interacting particles~\cite{Morr17Tin,Miya21NO2BHeavy} in recent years.

Even though impressive results have been obtained in nuclear many-body theory, several frontiers still remain:
\begin{itemize}
    \item[$i)$] the extension of \ai{} nuclear theory to heavy nuclei well above mass numbers $A \sim 100$,
    \item[$ii)$] many-body calculations for deformed nuclei that are not well approximated by a spherical reference state,
    \item[$iii)$] the systematic inclusion of higher-order terms in the many-body expansion for high-precision studies.
\end{itemize}
Some of the limiting factors common to all of these efforts are the computational and storage costs of the many-body calculation.
Calculations of larger systems require larger single-particle bases to converge calculations.
Similarly, calculations using symmetry-broken reference states (for instance, with axial symmetry rather than spherical symmetry) have to employ single-particles bases about an order of magnitude larger than standard spherically restricted calculations~\cite{Nova20PRCNeMgRch}.
Additionally, relaxing the many-body truncation in methods like the IMSRG and CC theory increases the scaling of both the computational and storage costs with respect to basis size,
and significant truncations have to be employed to make these calculations tractable~\cite{Nova20PRCNeMgRch,Hein21IMSRG3}.
Considering the computational challenges shared among these points, strategies to temper the storage and computational costs of the many-body expansion method would accelerate the progress for all of these developments.

It is well known that correlations are not uniformly distributed in the $A$-body Hilbert space and that certain configurations are more important for a quantitative reproduction of a given observable than others.
In a simplistic picture this is already implicitly used in the $\Xmax{N}$ truncation commonly employed in the no-core shell model (NCSM) that favors many-body states with low-lying excitations.
A more refined selection of the configuration space was first employed by the importance-truncated NCSM (IT-NCSM), which uses a perturbative selection measure to \textit{a priori} gauge the relevance of a given configuration for the final NCSM eigenstate.
With the aid of the IT it was first possible to extend the mass range of CI-based techniques to the oxygen drip line~\cite{Herg13MR}.
Similar ideas have since been employed in open-shell studies using particle-number-broken Hartree-Fock-Bogoliubov reference states.
While initially applied to Bogoliubov MBPT (BMBPT)~\cite{Tich19ITTF}, the IT concept was recently extended to nonperturbative Gorkov SCGF calculations~\cite{Porr21ITSCGF}.
More simplistic selection ideas based on natural orbital occupation numbers were used with great success in deformed CC applications~\cite{Nova20PRCNeMgRch}.

The goal of the present work is 
the extension of IT ideas to the IMSRG framework. 
We focus especially on the sensitivity of the importance-truncated results on the selection procedure and the interaction details for a wide range of IT measures.
Based on our findings we identify the IT-IMSRG as a promising candidate to reduce the costs of many-body calculations and, thus, extend the scope of \ai{} studies along the lines discussed above.

This paper is organized as follows.
In Sec.~\ref{sec:IMSRG} the IMSRG is introduced as the many-body formalism of choice.
Section~\ref{sec:IT} introduces the IT approach as applied to the IMSRG.
Results for selected mid-mass systems are presented in Sec.~\ref{sec:medium-mass_results}.
We conclude with a summary and perspectives in Sec.~\ref{sec:outlook}.

\section{In-medium similarity renormalization group}
\label{sec:IMSRG}

\subsection{Operator representation}
\label{sec:op_notation}

In the following, we employ operators in second-quantized form with the particular notation of the nuclear Hamiltonian up to the normal-ordered two-body level given by 
\begin{align}
    H = E_0 + \sum_{pq} f_{pq} :a^\dagger_p a_q: + \frac{1}{4} \sum_{pqrs} \Gamma_{pqrs} :a^\dagger_p a^\dagger_q a_s a_r: \,,
    \label{eq:hamiltonian_notation}
\end{align}
where $E_0$, $f_{pq}$, and $\Gamma_{pqrs}$ are the normal-ordered zero-, one\nobreakdash-, and (antisymmetrized) two-body matrix elements. Colons indicate strings of nucleon creation (annihilation) operators $a^{\dagger}_p$ ($a_p$) normal ordered with respect to a given $A$-body reference state $|\Phi \ra$. Consequently, the normal-ordered zero-body part is the reference-state expectation value, i.e., $E_0 = \la \Phi | H | \Phi \ra$.

The operator notation above employs a collective label for the single-particle state quantum numbers
\begin{align}
    | p \ra = | n l j m_{j}  t \ra \,,
\label{eq:collective_label}
\end{align}
where $n$ denotes the radial quantum number, $l$ the orbital angular momentum, $j$ the total angular momentum with projection $m_{j}$, and $t$ the isospin projection distinguishing proton and neutron states.
The model space in the following applications is constructed using the $\Xmax{e}$ truncation, which sets the largest principal quantum number $e \equiv 2n +l$ of states included in the single-particle Hilbert space $\mathcal{H}^{(1)}$. Typical values for converged calculations are around $\Xmax{e} \simeq 14$.
When including three-body operators, an additional restriction of the allowed three-body configurations $| p q r \ra$ to those with $e_p + e_q + e_r \leq E_{\text{3max}}$ is required to make calculations tractable.
Typical values are $E_{\text{3max}} \simeq 16$, but recent studies have extended this three-body truncation to substantially larger values of $E_{\text{3max}}=28$~\cite{Miya21NO2BHeavy}. Based on the representation in this basis space the 3\textit{N} interactions are then normal-ordered while all terms up to the two-body level are retained, as shown in Eq.~\eqref{eq:hamiltonian_notation}.

Various approaches exist to constructing the single-particle states in Eq.~\eqref{eq:collective_label}.
In this work, we use the perturbatively improved natural orbitals (NAT) basis~\cite{Tich19NatNCSM,Hopp2020natural,Nova20PRCNeMgRch}.
Within this basis, we construct a single Slater-determinant reference state for our many body calculations following the approach of Ref.~\cite{Hopp2020natural}.

\subsection{Many-body approach}
\label{sec:many_body_approach}

The IMSRG aims to decouple particle-hole excitations from the Slater-determinant reference state through a continuous unitary transformation $U(s)$ in the flow parameter $s$,
\begin{align}
    H(s)  = U(s) H U^{\dagger}(s) \, .\label{eq:imsrg_unitary}
\end{align}
The transformed Hamiltonian $H(s)$ is obtained by solving the flow equation 
\begin{align}
   \frac{d H(s)}{ds} = \left[ \eta(s), H(s) \right] \,,\label{eq:imsrg_diff_eq}
\end{align}
with the initial condition $H(0)=H$. 
The anti-Hermitian generator $\eta(s)$ is chosen to generate the desired decoupling behavior in the limit $s\rightarrow \infty$ when solving the system of coupled differential equations.
Once the decoupling is achieved, the Hamiltonian has effectively been diagonalized in the normal-ordered two-body approximation, and the correlated ground-state energy is obtained as the reference-state expectation value of the transformed Hamiltonian, i.e.,
\begin{align}
    E_\text{gs} = \lim_{s\rightarrow \infty} \la \Phi | H(s) | \Phi \ra \, .
\end{align}

The commutator evaluation in Eq.~\eqref{eq:imsrg_diff_eq} induces many-body operators, and so for practical applications a many-body truncation is necessary.
The standard approach is to truncate all operators at the normal-ordered two-body level, which gives the IMSRG(2) truncation.
In this work, we adapt the IMSRG(2) via importance truncation to give the IT-IMSRG(2),
using the Magnus expansion approach to solving the flow equation~\cite{Morr15Magnus}.
We refer to Refs.~\cite{Herg16PR,Herg17PS} for a detailed discussion of the many-body approach.

\subsection{Structure of two-body matrix elements}
\label{sec:2bme_structure}

Nuclear Hamiltonians obey a set of symmetries that one can explicitly exploit to lower computational requirements when storing operator matrix elements and performing many-body calculations.
These symmetries are rotational invariance ($[H,J^2]=[H,J_z]=0$), parity conservation ($[H,P]=0$), and isospin conservation ($[H,T_z]=0$).
Exploiting this leads to a block-diagonal structure, where we store the two-body matrix elements in separate ($JPT_z$) blocks.
This block structure is preserved for the normal-ordered Hamiltonian if a  symmetry-conserving reference state is employed,
which is the case for all systems and computational bases we consider.
Application of symmetry-broken reference states leads to many-body operators with lesser symmetries compared to their non-normal-ordered representations~\cite{Ripoche2020,Frosini2021}.

Many-body operators can be further decomposed in terms of their individual single-particle labels.
For a Slater-determinant reference state $|\Phi\ra$, one-body states [see Eq.~\eqref{eq:collective_label}] can be characterized by their occupation number $n_p$, where occupied ($n_p=1$) states are referred to as hole states and unoccupied ($n_p=0$) states as particle states. In the case of the two-body part of the operator, this leaves us with six interaction blocks of single-particle index combinations: hhhh, hhhp, hhpp, hphp, hppp, and pppp (plus their Hermitian conjugates and symmetry related blocks).
The notation we use here indicates that, e.g., for the hhpp block the two single-particle states in the bra two-body state are hole states and the two single-particle states in the ket two-body state are particle states.
In model-space sizes required for converged calculations, the number of particle states typically significantly exceeds the number of hole states.
Consequently, the pppp and hppp blocks drive the computational complexity of the many-body calculation, and blocks like the hhhh or hhhp blocks have a relatively small cost in terms of memory and computation.

We investigate the various two-body interaction blocks, their contributions to the perturbative energy corrections and different diagrams in the IMSRG, and their role in the IT-IMSRG in more detail in the next subsections.

\subsection{Perturbative analysis}
\label{sec:MBPT}

When following the standard Rayleigh-Schr\"odinger formulation of perturbation theory using the M{\o}ller-Plesset partitioning~\cite{Shav09MBmethod,Tichai2020review}, the canonical second-order (MP2) energy correction is given by
\begin{align}
    E^{(2)} &= \frac{1}{4} \sum \limits_{abij} \frac{\Gamma_{abij} \Gamma_{ijab}}{\varepsilon^{ab}_{ij}} \,,
\label{eq:MP2}
\end{align}
where the energy denominator 
$\varepsilon^{ab}_{ij} = \varepsilon_i + \varepsilon_j - \varepsilon_a - \varepsilon_b$ 
is given in terms of single-particle energies $\varepsilon_p = f_{pp}$ defined as the diagonal matrix elements of the normal-ordered one-body Hamiltonian.
Here and in the following, the indices $i$, $j$, $\ldots$ ($a$, $b$, $\ldots$) denote hole (particle) indices that are occupied (unoccupied) in the reference state, while $p$, $q$, $\ldots$ denote
generic single-particle indices.
From Eq.~\eqref{eq:MP2} we see that the second-order energy correction is only sensitive to the hhpp block of the interaction.

The third-order (MP3) energy correction in a canonical basis can be written as 
\begin{align}
    E^{(3)} = E^{(3)}_\text{pp} + E^{(3)}_\text{hh} + E^{(3)}_\text{ph} \, ,
\end{align}
where 
\begin{subequations}\label{eq:MP3}
\begin{align}
\label{eq:MP3_pp}
    E^{(3)}_\text{pp} &= \frac{1}{8} \sum \limits_{abcdij} \frac{\Gamma_{ijab} \Gamma_{abcd} \Gamma_{cdij}}{\varepsilon^{ab}_{ij} \varepsilon^{cd}_{ij}} \, ,\\
\label{eq:MP3_hh}    
    E^{(3)}_\text{hh} &= \frac{1}{8} \sum \limits_{abijkl} \frac{\Gamma_{ijab} \Gamma_{abkl} \Gamma_{klij}}{\varepsilon^{ab}_{ij} \varepsilon^{ab}_{kl}}\,, \\
\label{eq:MP3_ph}    
    E^{(3)}_\text{ph} &= - \sum \limits_{abcijk} \frac{\Gamma_{ijab} \Gamma_{kbic} \Gamma_{ackj}}{\varepsilon^{ab}_{ij} \varepsilon^{ac}_{kj}}
\end{align}
\end{subequations}
are the expressions for the pp (particle-particle), hh (hole-hole), and ph (particle-hole) diagrams, respectively.
Consequently, the third-order energy correction is sensitive to the hphp, pppp, and hhhh blocks in addition to the hhpp block.
Through a wide range of mass numbers it was shown that correlation effects from the particle-hole diagram dominate the third-order contribution for Hamiltonians that are amenable to MBPT~\cite{Tich16HFMBPT}.
This will become important later for selecting the most relevant subblocks for the preprocessing in the IT approach.

When working in a noncanonical basis, e.g., the NAT basis, the one-body part is not diagonal anymore and additional contributions to the perturbative energy corrections have to be considered that also include one-body vertices.
For the second-order energy correction, there is one additional diagram, and at third order 11 new diagrams arise~\cite{Shav09MBmethod}. The new third-order diagrams are now also sensitive to the two-body matrix elements in the hppp and hhhp blocks.
Consequently, all interaction blocks contribute to the energy correction up to third order in a noncanonical basis.

\section{Importance truncation}
\label{sec:IT}

\subsection{Rationale}
\label{sec:IT_rationale}

The aim of importance truncation is to effectively reduce the size of the problem by only considering the most important contributions based on a predefined importance measure.
Significant benefits are obtained by combining measures that are computationally cheap to construct with computationally more challenging many-body methods. 
Truncating unimportant parts of the many-body problem reduces the cost of the IT-adapted many-body method.
The discarded information can further be approximately accounted for in a perturbative way in order to minimize the information loss due to the IT.
Improving the quality of the IT measure or the approximation used to account for truncation effects can be used to reduce the systematic error introduced by the IT.
In practice, however, a reasonable balance between accuracy and complexity of the IT measure construction and the approximate treatment of IT-truncated parts has to be found. 

To understand how we approach IT in the IMSRG, it is instructive to review how importance truncation is performed in the IT-NSCM~\cite{Roth07ITCa40, Roth09ImTr}.
In the NCSM, the matrix elements of the Hamiltonian are evaluated in a basis of Slater determinants (configurations) $\{ | \Phi \ra \}$, and a subsequent diagonalization provides access to low-lying energies and corresponding eigenstates.
The IT approach is based on defining an importance measure $\kappa$ that gives an estimate of the importance of a particular configuration $| \Phi \ra$.
This measure is used to find the corresponding subspace of the $A$-body Hilbert space
\begin{align}
    \mathcal{M}^\text{IT}(\kappa_\text{min}) \equiv 
    \{ | \Phi \ra \, : \kappa(|\Phi\ra) \geq \kappa_{\min} \} 
    \subset \mathcal{H}^{(A)}
\end{align}
of the most important configurations with $\kappa(|\Phi\ra)$ above a chosen threshold $\kappa_{\text{min}}$.

The reduced size of the IT-selected subspace crucially reduces the computational cost of the following diagonalization, which roughly scales like the size of the subspace squared.
Residual effects from truncated configurations are approximately incorporated via a low-order multiconfiguration perturbation theory treatment~\cite{Rolik2003,Surjan2004}.
In the limit of $\Xmin{\kappa} \rightarrow 0$ the full configuration space is recovered and no approximation is induced by the importance truncation.

\subsection{Application to the IMSRG}

The IMSRG differs from the NCSM by employing a Fock space rather than a configuration space formulation.
As a result, the implementation of the IT approach needs to be adapted to work with many-body operators.
In the following, we focus the discussion of our approach on its application to two-body operators as in the IMSRG(2) they dominate the storage costs and contribute to the computationally dominating commutators.
The approach is, however, general and could be easily applied to one-body operators or three-body operators in IMSRG(3) calculations (with appropriately adapted importance measures).

To implement IT on two-body operator matrix elements,
we analyze whether the matrix element at the single-particle index combination $pqrs$ is important or not.
The result of this analysis is a ``mask'' based on our importance measure and chosen importance truncation threshold:
\begin{align}
    \kappa_{pqrs}^{\text{mask}} = 
    \begin{cases}
        1 & \text{if} \ pqrs \ \text{is important} \,, \\
        0 &  \text{otherwise} \,.
    \end{cases}
\end{align}
This is similar to how in the IT-NCSM configurations are analyzed and are either kept or removed (``masked'') from the model space.
The task is then to figure out which index combinations are ``important.''
We use importance measures $\kappa$
that take as input two-body matrix elements of some operator
(typically the Hamiltonian)
and give a value for the measure $\kappa_{pqrs}$ for each single-particle index combination.
If $\kappa_{pqrs} \geq \kappa_{\text{min}}$ for our chosen threshold $\kappa_{\text{min}}$,
we say that the index combination $pqrs$ is important,
and the mask $\kappa_{pqrs}^{\text{mask}}$ takes on a value of 1. In Sec.~\ref{sec:measures} we discuss different possible importance measures.

Given such a mask,
the matrix elements of a two-body operator $O_{pqrs}$
can be split into an important part,
\begin{align}
    O_{pqrs}^{\text{imp.}} = \kappa_{pqrs}^{\text{mask}} O_{pqrs}\,,
    \label{eq:it_important_matrix_elements}
\end{align}
and a residual part,
\begin{align}
\label{eq:IT_ME_res}
    O_{pqrs}^{\text{res.}} = (1 - \kappa_{pqrs}^{\text{mask}}) O_{pqrs}\,.
\end{align}
Keeping only the important part of all two-body matrix elements in the IMSRG amounts to solving the flow equation for only a subset of single-particle index combinations,
which gives us the IT-IMSRG.
The residual part may be treated approximately independent of the IT-IMSRG solution to capture the main effects neglected by its removal.

More concretely, in the IMSRG(2) we apply the IT approach discussed above to the two-body part of the Hamiltonian $\Gamma$, see Eq.~\eqref{eq:hamiltonian_notation}.
Additionally, the same truncation (i.e., using the same $\kappaX{\text{mask}}$) is applied to the two-body part of the generator and any two-body parts arising from commutator evaluations.
The initially removed matrix elements from the Hamiltonian will be treated perturbatively, and are not considered in the IMSRG solution.
Moreover, the evaluation of commutators in the IMSRG induces contributions to IT-neglected matrix elements in the resulting operator.
These are discarded and not treated further in our approach.

\begin{figure}
\centering
\includegraphics[width=\columnwidth,clip=]{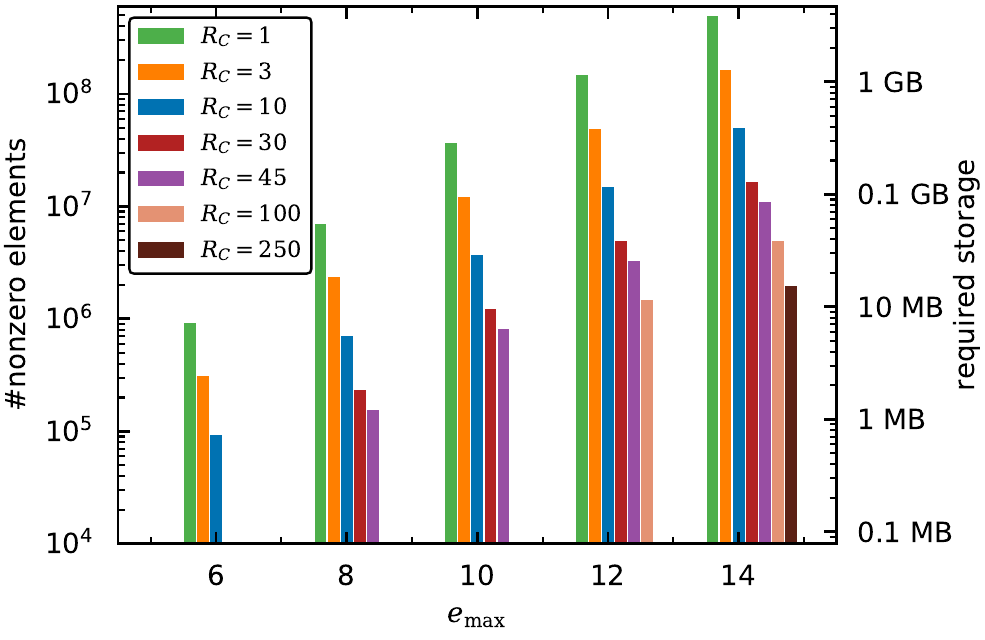}
\caption{
\label{fig:ms_compression}
Total number (left $y$ axis) and corresponding storage requirements (right $y$ axis) of nonzero two-body matrix elements for different compression ratios $R_C$ as a function of the model-space size $e_\text{max}$. For $R_C=1$ the number of matrix elements corresponds to the initial Hamiltonian without any IT.}
\end{figure}

The storage benefits of an IT-preprocessed operator are conveniently characterized by defining a compression ratio
\begin{align}
    R_C = \frac{\text{\# of nonzero MEs}}{\text{\# of nonzero MEs} - \text{\# of IT-neglected MEs}} \,,
\end{align}
given by the ratio of the number of initial nonzero two-body matrix elements (MEs) over the number of remaining nonzero two-body matrix elements after the IT.
In the case of no truncation $R_C=1$ and no compression is obtained.
Once the IT selection is performed, $R_C$ exceeds unity indicating a possibly lower memory footprint;
truncating 90\% of matrix elements gives $R_C=10$ and truncating 99\% of matrix elements gives $R_C=100$.
In this way the compression ratio provides an estimate of the scaling gained by the IT.
For example, a compression ratio of $R_C=70$ in an $e_\text{max}=14$ model space corresponds to truncating approximately 485.7 million of the total 492.7 million matrix elements.\footnote{
Displayed dimensionalities assume full exploitation of rotational invariance, parity and isospin conservation, and permutation symmetries.}
This leaves just 7 million nonvanishing matrix elements, which is roughly equivalent to an effective single-particle model space of $e_\text{max}=8$. 
This feature can also be clearly identified in the scaling example in Fig.~\ref{fig:ms_compression}, where we show the number of nonvanishing two-body matrix elements for different compression ratios $R_C$ obtained in different model-space sizes $e_\text{max}$.
In general, the application of IT techniques is expected to be most efficient in large model spaces since high-lying excitations typically contribute less to ground-state observables and, thus, larger compression ratios can be obtained without introducing significant errors with respect to the exact result.

\begin{figure}
\centering
\centering
\includegraphics[width=\columnwidth,clip=]{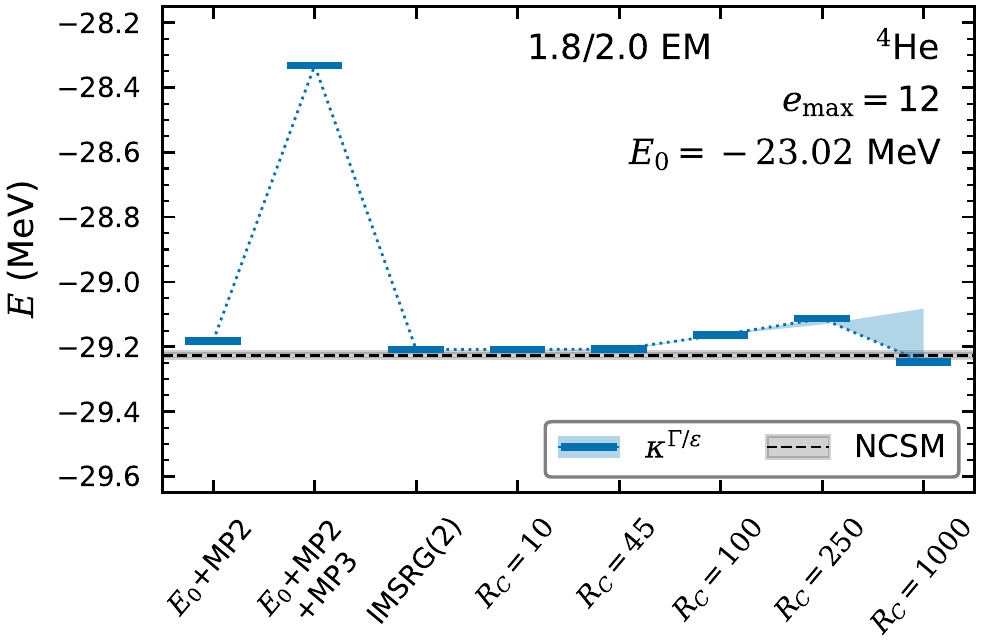}
\caption{
\label{fig:barplot_He4_18_20_EM}
Ground-state energy of $^{4}$He in the optimized $e^\text{NAT}=14$ NAT basis truncated to $e_\text{max}=12$ for different IT compression ratios $R_C$ and the IT measure $\kappa^{\Gamma/\varepsilon}$ (see Sec.~\ref{sec:measures}) compared to the full IMSRG(2) result and MBPT up to second and third order, using the 1.8/2.0 EM interaction. The third-order energy correction of the IT-neglected entries (as introduced in Sec.~\ref{sec:pert_correction_and_uncertainty}) is given by the (cyan) colored band. The black dashed line indicates the extrapolated NCSM result (see text for details).}
\end{figure}

A first example for what can be expected from the IT-IMSRG is shown in Fig.~\ref{fig:barplot_He4_18_20_EM}. While the specific details are explained in the following sections, the general trends we observe from IT-IMSRG(2) calculations at different $R_C$ are clear. 
We compare ground-state energies of $^4$He for a chosen IT measure, which is detailed in Sec.~\ref{sec:measures}, to the full IMSRG(2) result as well as to the second and third-order MBPT energy.
The colored band perturbatively incorporates the IT-neglected matrix elements, as outlined in Sec.~\ref{sec:pert_correction_and_uncertainty}.
We additionally show the extrapolated NCSM result for comparison, which is obtained by an extrapolation to $N_\text{max} \to \infty$ based on calculations up to $N_\text{max} = 14$ using the \texttt{BIGSTICK} code~\cite{John18Bigstick}.

\section{Medium-mass applications}
\label{sec:medium-mass_results}

In all following applications we use the IMSRG(2) solver by Stroberg~\cite{Stro17imsrggit}.
We note that the IT-IMSRG(2) implementation used in this work does not profit from the potential computational benefits of the IT framework by evaluating only important matrix-element index combinations in the flow.
Instead, we use the standard solver and set unimportant matrix elements to zero.
First computational studies for the advantages of an IT-IMSRG(2) solver using a sparse storage format are discussed in Sec.~\ref{sec:comput_benefits}.

\subsection{Interactions}
\label{sec:chiral_int}

In this work, we employ different \textit{NN} and 3\textit{N} interactions that are derived within the framework of chiral effective field theory (EFT).
We investigate two chiral interactions in detail: the next-to-next-to-next-to-leading order (N$^3$LO) \textit{NN} potential from Ref.~\cite{Ente17EMn4lo} with N$^3$LO 3\textit{N} forces constructed in Ref.~\cite{Dris17MCshort}, which in the following is referred to as ``N$^3$LO 500'' with cutoffs $\Lambda_{\textit{NN}} = \Lambda_{3\textit{N}} = 500$~MeV, and the ``1.8/2.0 EM'' interaction of Ref.~\cite{Hebe11fits} with N$^3$LO \textit{NN} and N$^2$LO 3\textit{N} forces.

Furthermore, we employ the free-space similarity renormalization group (SRG) (see Refs.~\cite{Bogn07SRG,Bogn10PPNP,Hebe203NF}) as a tool to evolve nuclear interactions to lower resolution scales, characterized by the flow parameter $\lambda$, by decoupling low- and high-momentum parts in the Hamiltonian.
Hamiltonians constructed using SRG-evolved potentials are perturbative, and their use in many-body applications leads to improved convergence behavior~\cite{Bogn10PPNP,Roth14SRG3N,Hopp19medmass,Hebe203NF}.
This is in contrast to the less-than-ideal convergence behavior of MBPT~\cite{Tich16HFMBPT}
and the possible appearance of unbound mean-field solutions (see, e.g., Ref.~\cite{Hopp2020natural}) when using harder unevolved potentials.
In this work, we study both the SRG-unevolved and consistently SRG-evolved \textit{NN}+3\textit{N} N$^3$LO 500 potential, where we refer to the unevolved potential if there is no additional SRG resolution scale given in the name specifier.
We note that for the unevolved N$^3$LO 500 potential the results are not converged in the shown model spaces.
The 1.8/2.0 EM potential is a low-resolution potential constructed via the SRG evolution of only the \textit{NN} force combined with a subsequent fit of the 3\textit{N} low-energy couplings to three- and four-body systems.

\begin{figure*}
\centering
\begin{minipage}[c]{.49\textwidth}
\centering
\includegraphics[width=\columnwidth,clip=]{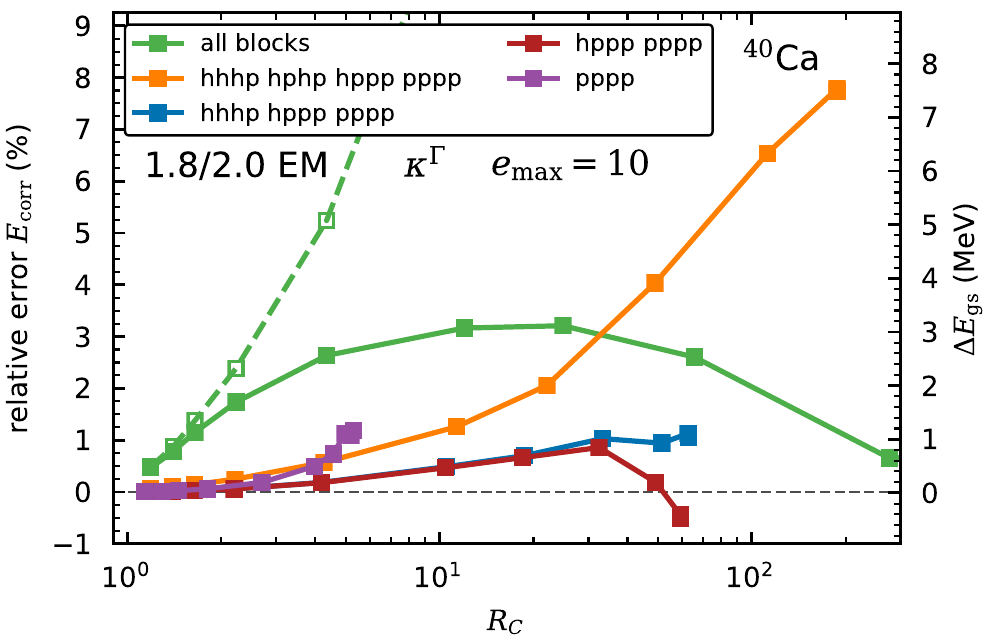}
\end{minipage}
\hspace*{0.16cm}
\begin{minipage}[c]{.49\textwidth}
\centering
\includegraphics[width=\columnwidth,clip=]{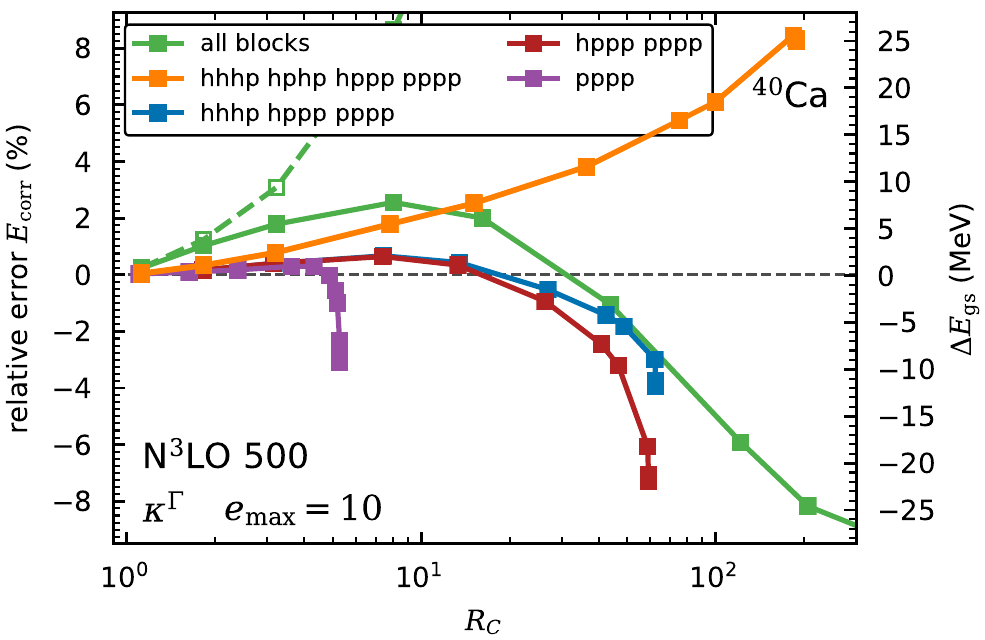}
\end{minipage}
\caption{
\label{fig:int_blocks_Ca40}
Error on the ground-state energy of $^{40}$Ca in the NAT basis as a function of compression ratio $R_C$ for IT-IMSRG(2) calculations truncating in various two-body interaction blocks.
The green curve for truncating in all two-body interaction blocks additionally incorporates the second-order energy correction of the neglected matrix elements. For comparison, we also show the results without the energy correction by the green dashed line with open squares. 
In the left panel the 1.8/2.0 EM Hamiltonian is used, and in the right panel the N$^3$LO 500 Hamiltonian is used.
The left $y$ axis indicates the relative error on the correlation energy $E_{\text{corr}}$, and the right $y$ axis shows the absolute error on the ground-state energy. For comparison the correlation energy is given by $E_\text{corr} = -96.88$~MeV and $E_\text{corr} = -303.02$~MeV for the left and right panels, respectively. 
All calculations are performed in an $\Xmax{e}=10$ model space using $\kappaX{\Gamma}$ (see Sec.~\ref{sec:measures} for details).}
\end{figure*}

\subsection{Interaction blocks}
\label{sec:int_blocks}

Before introducing the individual IT measures and investigating them in detail, we perform a more careful analysis of how sensitive the IT-IMSRG(2) solution is to truncations in the different two-body interaction blocks.
In Fig.~\ref{fig:int_blocks_Ca40}, we consider the IT-IMSRG(2) solution of ${}^{40}\text{Ca}$ when different combinations of blocks are truncated (indicated by the different lines) using the IT measure $\kappaX{\Gamma}$ based on the magnitude of the two-body matrix elements, which is introduced in Sec.~\ref{sec:measures}.
For each line, each point corresponds to a chosen $\kappa_{\text{min}}$, which gives a compression ratio $R_C$ and produces an error to the exact IMSRG(2) solution.
This error is shown in terms of the relative error on the correlation energy $E_{\text{corr}} = E(s\rightarrow \infty) - E(s=0)$ on the left $y$ axis
and in terms of the absolute error on the right $y$ axis.
The two panels differ in the interaction used: the left panel features the 1.8/2.0 EM Hamiltonian and the right panel the N$^3$LO 500 Hamiltonian.
In this figure, the truncated residual part of the two-body Hamiltonian is not treated approximately except when all blocks are truncated.
In this case, the second-order energy correction based on the hhpp block of the residual part is added to the IT-IMSRG(2) result.

In both panels, truncating all interaction blocks (the green curve) leads to the largest error at small and intermediate compression ratios.
Fully truncating the entire two-body part of the Hamiltonian causes the IT-IMSRG(2) to produce a ground-state energy that is exactly $E_0 + E^{(2)}$ thanks to the perturbative treatment of the truncated hhpp part of the Hamiltonian.
In the left panel this is remarkably close to the full IMSRG(2) result, but in the right panel the relative error to the full IMSRG(2) correlation energy is nearly 10\%.
This behavior can be systematically improved by restricting importance truncation to selected blocks.

The first blocks we remove from the importance truncation are the hhpp and hhhh blocks, which gives the orange curves in Fig.~\ref{fig:int_blocks_Ca40}.
We observed that truncating the hhpp block leads to large errors at intermediate compression ratios, motivating its exclusion from the IT.
The hhpp block [and the hhhppp block in the IMSRG(3)] have previously been observed to be quite important for the IMSRG~\cite{Herg16PR,Hein21IMSRG3}, and this finding supports that intuition further.
The hhhh block is the smallest block and, as a result, does not offer much room for compression, so we also leave it untruncated.
We find that the removal of these two blocks from the IT reduces the error at intermediate compressions substantially.
Still, the error grows large as we approach the maximum compression ratio.

Additionally removing the hphp block from the truncation yields the blue curve, which offers a substantial reduction in the error relative to the orange curve at all compression ratios.
As a reminder, the hphp block contributes at third order in MBPT (see Sec.~\ref{sec:MBPT}),
and in MBPT studies it was found that the third-order particle-hole diagram sensitive to this block dominates the third-order contribution~\cite{Tich16HFMBPT}.
In the IMSRG, it has been observed that the high-order generalizations of the particle-hole diagram, the ph-ring diagrams, which are resummed nonperturbatively in the IMSRG(2), are particularly important~\cite{Herg16PR,Stroberg2019}.
The substantial reduction in the IT-induced error we find when the hphp block is excluded is consistent with these observations.
Further restricting the IT to only the hppp and pppp blocks (leaving the small hhhp block untruncated) gives the red line,
and truncating only in the pppp block gives the purple line.
In our studies, we found that restricting the IT to just the hppp and pppp blocks (i.e., the red curves in Fig.~\ref{fig:int_blocks_Ca40}) allowed us to achieve large compression ratios while introducing relatively small errors for appropriate IT thresholds,
hence in the following we concentrate on this approach.

\subsection{Definition of importance measures}
\label{sec:measures}

In the following, various importance measures for the flowing two-body part of the Hamiltonian in the IMSRG(2) approach are investigated.
We note that in principle similar studies can be performed for the one-body part as well.
However, since the computational gain is negligible
we will focus on the two-body part here. All measures are constructed once at the beginning of the flow based on the initial Hamiltonian at $s=0$. Important combinations of single-particle indices $p$, $q$, $r$, and $s$ of $\Gamma_{pqrs}$ are identified by the IT measure and kept over the course of the flow, while matrix elements with unimportant index combinations are set to zero throughout the flow. Note that this also includes matrix elements which are potentially induced by the IMSRG.
An alternative approach would be to dynamically update the measure during the IMSRG evolution, which could possibly better account for the changing structure of the evolving Hamiltonian.
We leave the exploration of such strategies to future studies and focus instead on different measure choices and their relative performance for different systems.

\subsubsection{Matrix-element-based measures}

The simplest way of estimating the relevance of a given two-body matrix element is its initial magnitude, giving rise to the first importance measure,
\begin{align}
    \label{eq:ITmeasures_GammaMin}
     \kappaX{\Gamma}_{pqrs}(\Gamma)& = \Big | \Gamma_{pqrs} \Big | \, .
\end{align}
This measure encodes the expectation that the largest matrix elements are expected to be most important for the evolution of the Hamiltonian and the smallest ones will be relatively unimportant.
However, this naive estimate does not take into account specific information about the target nucleus.

A more refined measure can be obtained by taking inspiration from MBPT.
In MBPT, the matrix elements that are summed over are accompanied by energy denominators,
and our measure based on this idea is
\begin{align}
    \kappaX{\Gamma/\varepsilon}_{pqrs}(\Gamma) = \Big | \Gamma_{pqrs}/\varepsilon_\text{sum} \Big | \, .
    \label{eq:ITmeasures_TBME_SPE}    
\end{align}
The additional appearance of an energy denominator $\varepsilon_\text{sum}$ accounts for the lower importance of highly excited configurations associated with large single-particle energies.
The measure $\kappaX{\Gamma/\varepsilon}$ is very closely related to the first-order estimate of the MBPT wave function expansion and similar in spirit to previously used measures in importance-truncated NCSM~\cite{Roth09ImTr}, BMBPT~\cite{Tich19ITTF}, or Gorkov SCGF~\cite{Porr21ITSCGF} frameworks.

The energy denominator for the hhpp block is simply given by $\varepsilon^{pp}_{hh}$ used in the second-order MBPT energy correction in Eq.~\eqref{eq:MP2}.
However, in the IMSRG other interaction blocks than only the hhpp block are present and the energy denominator of the importance measure has to be generalized accordingly.
This is done by defining 
\begin{align}
\label{eq:ITmeasure_SPE}
    \varepsilon_\text{sum} = 
    \begin{cases}
        4 \varepsilon_\text{F} - \varepsilon_{p_1} - \varepsilon_{p_2} - \varepsilon_{p_3} - \varepsilon_{p_4}  & \text{for pppp,} \\ \\
        2 \varepsilon_\text{F} + \varepsilon_{h}   - \varepsilon_{p_1} - \varepsilon_{p_2} - \varepsilon_{p_3}    & \text{for hppp,} \\
    \end{cases}
\end{align}
where $\varepsilon_\text{F}$ is the Fermi energy, i.e., the energy of the energetically highest-lying hole orbital.
Generally, the Fermi energy $\varepsilon_\text{F}$ is different for protons and neutrons in proton-rich or neutron-rich systems. While in this work we show results for the simple definition of an isospin-independent Fermi energy given above, we have explored using isospin-differentiated Fermi energies
and found that the different approaches give quantitatively very similar results in a broad range of systems.
Equation~\eqref{eq:ITmeasure_SPE} can be trivially extended to include the hphp, hhhp, and hhhh blocks, but we focus our discussion in this work on truncations of the pppp and hppp blocks, as explained in Sec.~\ref{sec:int_blocks}.

\subsubsection{Occupation-based measures}

In the natural orbitals basis, additional information about the system is available in the form of the noninteger occupation numbers $n^\text{NAT} \in [0,1]$ of the individual orbitals.
We can use this information to construct alternative truncation measures that only work when such noninteger occupation numbers, resulting from an improvement of the one-body density matrix beyond the mean-field level, are available.
The simplest choice is the use of products of occupation numbers obtained in the diagonalization of the one-body density matrix, as used in CC applications to improve convergence in the triples amplitudes truncation~\cite{Nova20PRCNeMgRch},
\begin{align}
    \kappaX{n}_{pqrs} =
    \prod_{i \in \{p, q, r, s\}}
    \begin{cases}
        | n^{\text{NAT}}_i| & \text{if $i$ is a particle state,}  \\ \\
        |1 - n^{\text{NAT}}_i| & \text{if $i$ is a hole state,}
    \end{cases}
    \label{eq:ITmeasures_NAT}    
\end{align}
where the product is given by the natural orbital occupation numbers $n^\text{NAT}$ for the $p$, $q$, $r$, and $s$ orbitals, with $n_i^\text{NAT}$ for particle states and $\overline{n}_i^\text{NAT} = (1-n_i^\text{NAT})$ for hole states.
This measure gives the greatest importance to matrix elements where bra and ket indices lie close to the Fermi surface.
This reflects the intuition that for low-resolution Hamiltonians the correlation expansion is dominated by low-energy excitations around the Fermi surface.
Again, this measure is rather simplistic since no explicit information from the two-body matrix elements enters the IT selection.

A further refinement is obtained by accounting for the magnitude of the associated two-body matrix element as is done in Eqs.~\eqref{eq:ITmeasures_GammaMin} and ~\eqref{eq:ITmeasures_TBME_SPE} via
\begin{align}
    \kappaX{\Gamma n}_{pqrs} (\Gamma) = \Big|\Gamma_{pqrs} \Big| \times
     \prod_{i \in \{p, q, r, s\}}
    \begin{cases}
        | n^{\text{NAT}}_i| & \text{if $i$ is a particle state,}  \\ \\
        |1 - n^{\text{NAT}}_i| & \text{if $i$ is a hole state.}
    \end{cases} 
    \label{eq:ITmeasures_TBME_NAT} 
\end{align}
In both cases,
the natural orbital occupation numbers contain additional information about the shell structure, such that contributions from, e.g., high orbital angular momentum (large $l$) or high radial excitations (large $n$) will typically be suppressed.

\begin{figure*}
\centering
\begin{minipage}[c]{.49\textwidth}
\centering
\includegraphics[width=\columnwidth,clip=]{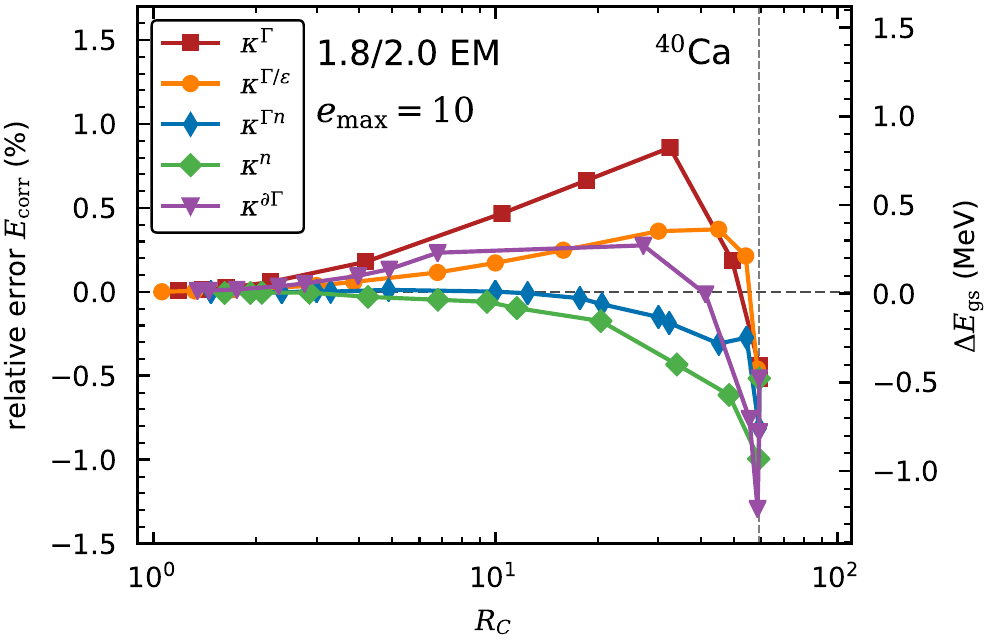}
\end{minipage}
\hspace*{0.16cm}
\begin{minipage}[c]{.49\textwidth}
\centering
\includegraphics[width=\columnwidth,clip=]{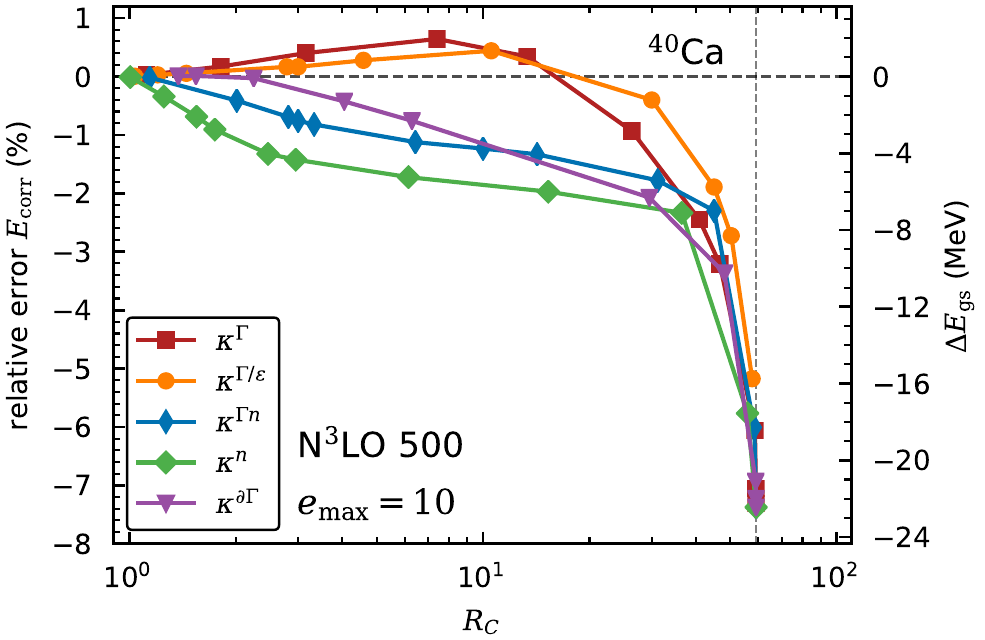}
\end{minipage}
\caption{
\label{fig:IT_measures_Ca40_e8}
Error on the ground-state energy of $^{40}$Ca in the NAT basis as a function of compression ratio for IT-IMSRG(2) calculations using five IT measures defined in Eqs.~\eqref{eq:ITmeasures_GammaMin}-\eqref{eq:ITmeasures_derivGamma}.
In the left panel the 1.8/2.0 EM Hamiltonian is used, and in the right panel the N$^3$LO 500 Hamiltonian is used.
The left $y$ axis indicates the relative error on the correlation energy $E_{\text{corr}}$, and the right $y$ axis shows the absolute error on the ground-state energy.
All calculations are performed in an $\Xmax{e}=10$ model space and the dashed vertical line indicates the maximum compression ratio.
}
\end{figure*}

\subsubsection{Derivative-based measures}

While the matrix-element- and occupation-based measures above are inspired by other many-body frameworks, the notion of derivative-based measures is specific to the IMSRG approach.
By defining the IT measure as
\begin{align}
    \kappaX{\partial \Gamma}_{pqrs} (\Gamma)
    = \Big |\left(\frac{dH}{ds}\right)^{(2)}_{pqrs} \Big|
    = \Big |\left[\eta,H\right]^{(2)}_{pqrs} \Big| \, ,
    \label{eq:ITmeasures_derivGamma}    
\end{align}
the importance of two-body matrix elements is based on the magnitude of their expected change.
Matrix elements with a large derivative are expected to change significantly over the course of the evolution, and the initial value will be a poor approximation.
Note, however, that this measure does not directly account for the size of the matrix element but only its expected dynamics independent of the starting value.
It is also worth mentioning that the construction of this measure is more expensive than the previously discussed measures, as the evaluation of the required commutator scales like $\mathcal{O}(N^6)$ in the size of the single-particle basis.

\subsection{Perturbative treatment of truncated Hamiltonian}
\label{sec:pert_correction_and_uncertainty}

In order to perturbatively consider the IT-neglected contributions, we apply a modified version of the MP3 energy correction with all diagrams sensitive to the pppp and hppp truncated interaction blocks. 
We consider the pp-ladder diagram $E^{(3)}_\text{pp}$ shown in Eq.~\eqref{eq:MP3_pp}, which is sensitive to the pppp matrix elements, as well as the two noncanonical diagrams that are sensitive to the hppp matrix elements, when working in a natural orbital basis. 
Combining the IT matrix elements of the two relevant interaction blocks with the initial hhpp elements results in an adapted third-order IT energy correction of the neglected contributions
\begin{align}
    E^{(3)}_\text{IT} = E^{(3)}_\text{pp} + E^{(3)}_\text{hppp} \,,
\end{align}
with the two parts given by 
\begin{align}
    E^{(3)}_\text{pp} &= \frac{1}{8} \sum \limits_{abcdij} \frac{\Gamma_{ijab} \ \Gamma^\text{res.}_{abcd} \ \Gamma_{cdij}}{\varepsilon^{ab}_{ij} \varepsilon^{cd}_{ij}} \,, \\
    E^{(3)}_\text{hppp} &= \frac{1}{2} \sum \limits_{abcij} \frac{\Gamma_{ijab} \ \Gamma^\text{res.}_{abcj} \ f_{ci}}{\varepsilon^{ab}_{ij}\varepsilon^c_i}  + \frac{1}{2} \sum \limits_{abcij} \frac{f_{ai} \ \Gamma^\text{res.}_{ajcb} \ \Gamma_{cbij}}{\varepsilon^a_i \varepsilon^{bc}_{ij}} \,, 
\end{align}
where the IT-neglected matrix elements are given by $\Gamma^\text{res.}$ as defined in Eq.~\eqref{eq:IT_ME_res} and the matrix elements $\Gamma$ and $f$ without superscript correspond to the initial two- and one-body contributions of the normal-ordered Hamiltonian in Eq.~\eqref{eq:hamiltonian_notation}.

\subsection{Measure sensitivity and compression benchmark}

The five different IT measures $\kappa$ introduced in Sec.~\ref{sec:measures} are studied for $^{40}$Ca using two different Hamiltonians in Fig.~\ref{fig:IT_measures_Ca40_e8}.
For both Hamiltonians one finds that at the maximum compression accessible for the chosen interaction blocks (indicated by the dashed vertical line) all measures give the same error to the exact IMSRG(2) result, which reflects that at this compression $\kappa_{\text{min}}$ for each measure has been chosen such that the hppp and pppp blocks are completely truncated.
At intermediate compressions, however, the various measures give different results.
Of particular interest is the growth in the error to the exact IMSRG(2) result when going from small to intermediate compressions.
Concentrating first on the matrix-element-based measures (the red squares and orange circles), we find that they follow the same qualitative trend.
The more refined $\kappaX{\Gamma/\varepsilon}$ gives smaller errors than $\kappaX{\Gamma}$ at small and intermediate compressions,
especially for the 1.8/2.0 Hamiltonian.
The fact that $\kappaX{\Gamma/\varepsilon}$ works so well for the 1.8/2.0 Hamiltonian reflects the perturbativeness of the Hamiltonian, but the MBPT-inspired refinement over $\kappaX{\Gamma}$ seems to be effective for harder interactions as well.
\begin{figure}[t!]
\centering
\includegraphics[width=\columnwidth,clip=]{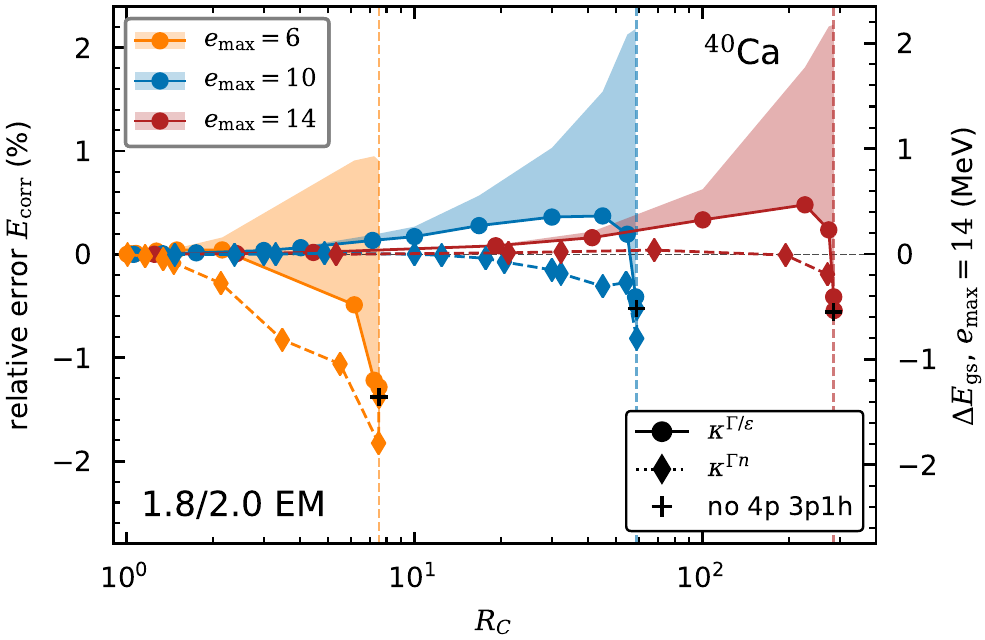}
\caption{
\label{fig:modelspace_Ca40}
Relative error of the correlation energy in the NAT basis of $^{40}$Ca using the 1.8/2.0 EM Hamiltonian as a function of the compression ratio for the model-space sizes $e_\text{max}=6$, 10, and 14.
We show results for the IT measures $\kappaX{\Gamma/\varepsilon}$ (circles) and $\kappaX{\Gamma n}$ (diamonds).
The right $y$ axis indicates the absolute difference between the $e_\text{max}=14$ IT-IMSRG(2) results and the exact IMSRG(2) result.
The MP3 energy correction for the $\kappaX{\Gamma/\varepsilon}$ IT-neglected contributions is indicated by the corresponding band for each model-space truncation and the vertical dashed lines correspond to the maximum $R_C$ for the given model space.
}
\end{figure}

Turning to the occupation-based measures (the green diamonds and the blue thin diamonds),
we find that they produce smaller errors than $\kappaX{\Gamma/\varepsilon}$ and $\kappaX{\Gamma}$ for the 1.8/2.0 EM Hamiltonian and larger errors for the N$^3$LO 500 Hamiltonian.
This can also be understood due to the relative softness of the two Hamiltonians, as the natural orbitals are constructed using second-order MBPT.
It is likely that for harder Hamiltonians this construction does not approximate the one-body density matrix well enough for the IT measure based on its occupation numbers to be effective.
We also find that refining $\kappaX{n}$ by including the matrix element size to give $\kappaX{\Gamma n}$ produces smaller errors at all compression ratios.
We find that the derivative-based measure $\kappaX{\partial \Gamma}$ (purple triangles) performs similarly to the other measures investigated, but costs substantially more to construct.

As outlined in Sec.~\ref{sec:2bme_structure}, we utilize the underlying symmetry of the Hamiltonian to store the two-body matrix elements in a $(JPT_z)$ block structure and apply the IT in the individual symmetry blocks.
Although the resulting energy can be quite different depending on the chosen IT measure, 
we observe nearly identical suppression of the number of matrix elements in the $(JPT_z)$ blocks for the three IT measures $\kappa^\Gamma$, $\kappa^{\Gamma/\varepsilon}$, and $\kappa^{\Gamma n}$ at comparable compression ratios.
In the rest of this work, we focus our explorations mostly on the two best performing measures, $\kappaX{\Gamma/\varepsilon}$ and $\kappaX{\Gamma n}$ [see Eqs.~\eqref{eq:ITmeasures_TBME_SPE} and~\eqref{eq:ITmeasures_TBME_NAT}].

In Fig.~\ref{fig:modelspace_Ca40},
we study the effect of going to different model-space sizes on the achievable compression ratios and their associated errors in $^{40}$Ca with the 1.8/2.0 EM Hamiltonian.
Going from $e_{\text{max}}=10$ to $e_{\text{max}}=14$ increases the number of two-body matrix elements by a factor of roughly 15,
which allows for higher maximum compression.
However, at the same compression ratio, the $e_{\text{max}}=14$ IT-IMSRG(2) calculations have much smaller errors to the exact result, because many of the matrix elements that are added when going from $e_{\text{max}}=10$ to $e_{\text{max}}=14$ can be truncated.
This also means that the IT-IMSRG(2) is less effective in small model spaces (such as $e_{\text{max}}=6$ in Fig.~\ref{fig:modelspace_Ca40}) because in these model spaces the maximum compression is lower and truncating a lot of matrix elements quickly leads to larger errors.

In Fig.~\ref{fig:modelspace_Ca40},
we include the perturbative treatment of the truncated part of the Hamiltonian introduced in Sec.~\ref{sec:pert_correction_and_uncertainty}.
The points on the lines are the result of IT-IMSRG(2) calculations without any extra treatment of the truncated part.
The MP3 correction due to the IT-neglected matrix elements is included as a band on top of this IT-IMSRG(2) result.
We understand this correction to be an indication of the magnitude of the missing third-order and higher-order contributions that are discarded by the truncation.
This can be used as an estimate of the IT uncertainty relative to the full IMSRG(2) for a given truncation threshold,
which is important in cases where exact results are not readily available for comparison.
We see that in this case the uncertainty indicated by the MP3 correction is reasonable up to relatively high compression ratios.
As an example, for an $e_{\text{max}}=14$ IT-IMSRG(2) calculation, we can achieve a compression ratio of 100 while keeping the IT uncertainty below 0.5\% on the correlation energy.
\begin{figure}[t!]
\centering
\includegraphics[width=\columnwidth,clip=]{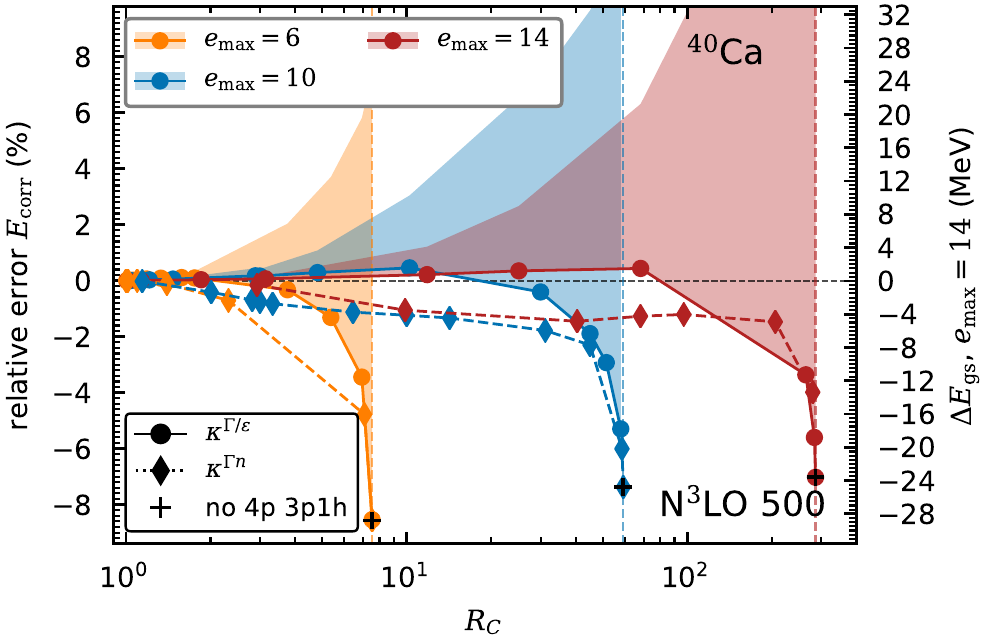}
\caption{
\label{fig:modelspace_Ca40_EMN}
Same as Fig.~\ref{fig:modelspace_Ca40} but for the N$^3$LO 500 Hamiltonian.}
\end{figure}

\begin{figure}[b!]
\centering
\includegraphics[width=\columnwidth,clip=]{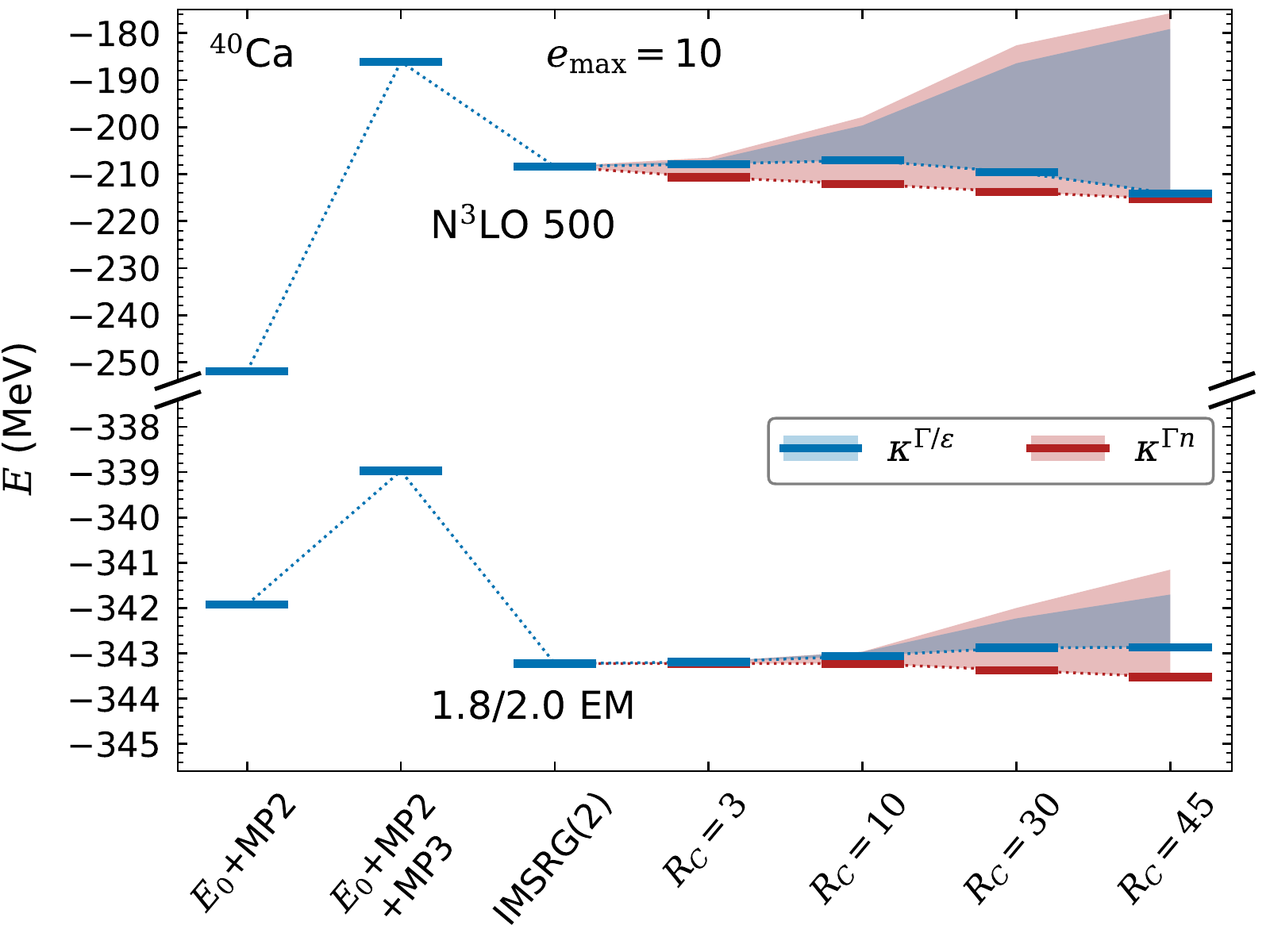}
\caption{
\label{fig:barplot_Ca40_e10}
Ground-state energy of $^{40}$Ca in the NAT basis for model-space size $e_\text{max}=10$ and two IT measures as a function of compression ratios $R_C$ compared to the full IMSRG(2) result and results from second- and third-order MBPT, using the N$^3$LO 500 interaction (top) and the 1.8/2.0 EM interaction (bottom). The third-order energy correction of the IT-neglected contributions for both measures is given by the correspondingly colored band.}
\end{figure}

\subsection{Interaction sensitivity}

We now turn our attention to how the observed trends are affected by the choice of Hamiltonian.
In Fig.~\ref{fig:modelspace_Ca40_EMN},
we again show the IT-IMSRG(2) errors in several model-space sizes for $^{40}$Ca, this time using the N$^3$LO 500 Hamiltonian.
We see that, compared to the softer 1.8/2.0 EM Hamiltonian, the relative error on the correlation energy is similar, which is due to the large correlation energy for the N$^3$LO 500 Hamiltonian.
Accordingly, the absolute errors in the energy are larger than in Fig.~\ref{fig:modelspace_Ca40}.
We see that the IT-IMSRG(2) results based on $\kappaX{\Gamma/\varepsilon}$ are better than using $\kappaX{\Gamma n}$,
and the uncorrected results for $\kappaX{\Gamma/\varepsilon}$ lie very close to the exact IMSRG(2) results up to compressions near 100 for $e_{\text{max}}=14$.
However, the MP3 correction due to the IT-neglected matrix elements is much larger for the harder N$^3$LO 500 Hamiltonian, suggesting a sizable uncertainty in the IT-IMSRG(2) results even at relatively small compression ratios.
\begin{figure}[t!]
\centering
\includegraphics[width=\columnwidth,clip=]{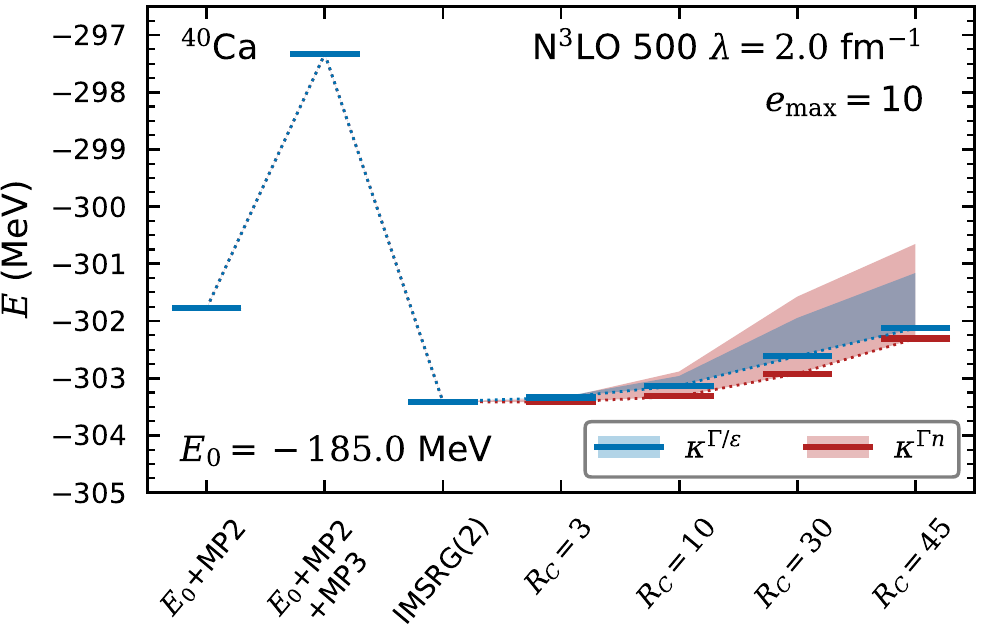}
\caption{
\label{fig:barplot_Ca40_e10_EMN_2.0}
Same as Fig.~\ref{fig:barplot_Ca40_e10} but for the SRG-evolved N$^3$LO 500 interaction to $\lambda=2.0$~fm$^{-1}$.}
\end{figure}

In Fig.~\ref{fig:barplot_Ca40_e10},
we compare ground-state energies for $^{40}$Ca obtained via various different many-body approaches for the N$^3$LO 500 and 1.8/2.0 EM Hamiltonians in an $e_{\text{max}}=10$ model space.
We compare IT-IMSRG(2) results at different compression ratios against untruncated IMSRG(2) results
and provide results from second- and third-order MBPT for comparison.
Looking first at the results for the 1.8/2.0 Hamiltonian,
we see that second- and third-order MBPT energies differ from the IMSRG(2) result by less than 5~MeV.
Overall, the IT-IMSRG(2) results also agree well with the IMSRG(2) results. In particular, for intermediate compression ratios ($R_C=10$, for instance) the estimated uncertainty from the treatment of IT-neglected matrix elements is very small.
For the N$^3$LO 500 Hamiltonian, this picture changes substantially.
The second-order MBPT, third-order MBPT, and IMSRG(2) results span an energy range of about 65~MeV.
Here the IT-IMSRG(2) performs quite well, giving errors to the IMSRG(2) of up to roughly 7~MeV for the largest compression ratio considered in Fig.~\ref{fig:barplot_Ca40_e10}.
The estimated uncertainty, however, is much larger than the actually observed errors.
It seems that the correction we obtain from the IT-neglected matrix elements for this interaction overestimates the size of the missing physics,
which we understand to be an artifact of the perturbative approach we take in conjunction with a nonperturbative Hamiltonian.

To confirm this,
we consider the same system using the N$^3$LO 500 Hamiltonian consistently SRG evolved to a resolution scale of $\lambda=2.0$~fm$^{-1}$ in Fig.~\ref{fig:barplot_Ca40_e10_EMN_2.0}.
This SRG-evolved potential is very soft,
as also suggested by the small differences between second-order MBPT, third-order MBPT, and IMSRG(2) results.
Here the difference between the IT-IMSRG(2) and the IMSRG(2) results is about 1.5~MeV for $R_C=45$,
and the estimated uncertainty is also of approximately the same size as the error.
This suggests that our uncertainty estimate is quite reasonable for soft Hamiltonians and the IT results for an intermediate compression ratio of about $R_C=10$ lie very close to the full IMSRG(2) result for such interactions.

\subsection{Mass number sensitivity}

\begin{figure}[t]
\centering
\includegraphics[width=\columnwidth,clip=]{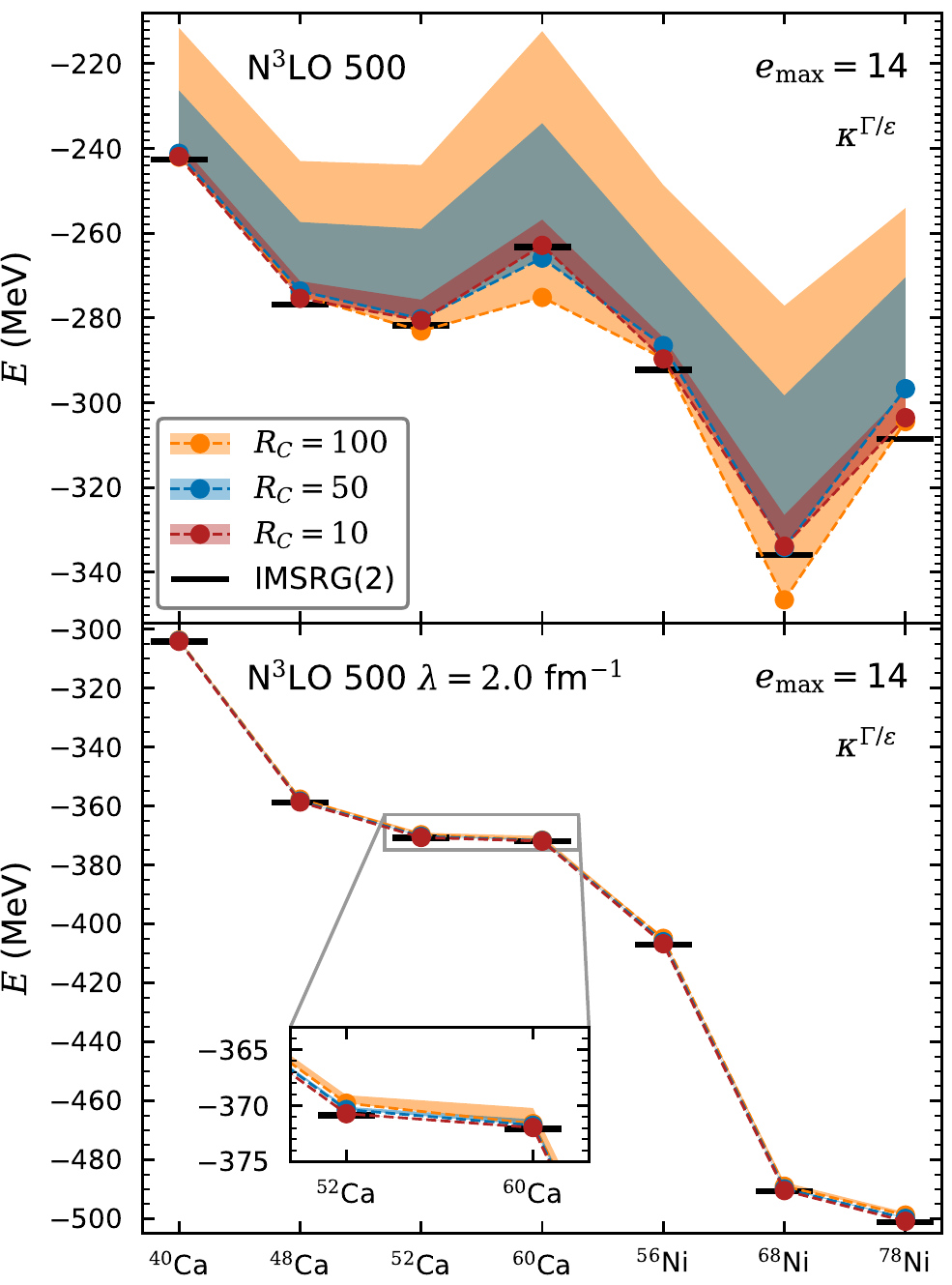}
\caption{
\label{fig:med_mass_CaNi_EMN_10}
Ground-state energies for selected calcium and nickel nuclei in the IT-IMSRG with compression ratios $R_C= 10$, 50, and 100 for the measure $\kappa^{\Gamma / \varepsilon}$. Results are shown for the NAT basis and the unevolved (top) and SRG-evolved to $\lambda=2.0$~fm$^{-1}$ (bottom) N$^3$LO 500 interaction in $e_\text{max}=14$.}
\end{figure}

After our investigation of the interaction blocks and IT measures for $^{40}$Ca as a reasonable test case, we extend our studies to closed-shell nuclei ranging from $^{40}$Ca up to $^{78}$Ni.
In the top panel of Fig.~\ref{fig:med_mass_CaNi_EMN_10}, we show ground-state energies for different compression ratios in the IT-IMSRG(2) compared to the full IMSRG(2) result when using the N$^3$LO 500 Hamiltonian.
At intermediate compressions (here $R_C=10$), we find that the IT-IMSRG(2) results match the exact IMSRG(2) results within a few MeV.
The largest deviation is found for $^{78}$Ni with an error of 5~MeV.
The uncertainty indicated by our third-order treatment of truncated matrix elements is also of about the same size.
Going to larger compressions, we find that at $R_C=50$ the IT-IMSRG(2) results still lie remarkably close to the exact IMSRG(2) results.
The 12~MeV error for $^{78}$Ni is the exception here, and most errors are still below 5~MeV.
For $R_C=100$, the deviation to exact IMSRG(2) results tends to be larger.
However, for both $R_C=50$ and $R_C=100$ the indicated uncertainty is much larger than the observed error, growing to beyond 50~MeV in some systems.
We emphasize that for these results $R_C=50$ and $R_C=100$ are being used to explore the truncation error for heavily truncated calculations
and would not be considered adequate for practical nuclear structure calculations.

We perform the same study across a variety of systems in the lower panel of Fig.~\ref{fig:med_mass_CaNi_EMN_10}, but this time using the SRG-evolved N$^3$LO 500 $\lambda=2.0$~fm$^{-1}$ Hamiltonian, which in the last section showed substantially smaller errors in IT-IMSRG(2) calculations than the unevolved N$^3$LO 500 Hamiltonian.
For this Hamiltonian, we see that even with $R_C=100$ the error to the exact IMSRG(2) result is very small in all systems (with a maximum of 2.5~MeV for $^{78}$Ni).
Moreover, the uncertainty indicated by the third-order treatment of truncated matrix elements is also quite small, only growing up to 1.4~MeV for $^{68}$Ni with $R_C=100$.
These results show the promising performance of the IT-IMSRG over a wide range of mass numbers and up to large compression ratios when using softer or SRG-evolved interactions.
While Fig.~\ref{fig:med_mass_CaNi_EMN_10} only shows results obtained using the $\kappaX{\Gamma/\varepsilon}$ measure, the results are very similar for $\kappaX{\Gamma n}$.

\subsection{First computational investigations}
\label{sec:comput_benefits}

In this work, we have introduced the IT-IMSRG formalism as a way to approximate IMSRG(2) results while considering only a small subset of the two-body matrix elements.
In our implementation, we applied the formalism by taking the IMSRG solver by Stroberg~\cite{Stro17imsrggit} and explicitly setting truncated matrix elements to zero.
An implementation that fully takes advantage of the new structure of importance-truncated operators is beyond the scope of this work, but will be necessary to take advantage of the storage and computational benefits of the IT-IMSRG.
We performed an initial exploration of this by adapting the solver mentioned above to use a modified storage format.

An optimized IT-IMSRG solver must not store the zeros associated with truncated matrix elements,
and it must be able to do so flexibly as different IT measures will truncate different matrix elements.
This naturally suggests the use of sparse matrices and sparse linear algebra operations (as is available in, e.g., the \texttt{C++} \textsc{armadillo} library~\cite{Sand16armadillo,Sand18armaSP}).
We were able to adapt the storage format of our IT-IMSRG(2) solver to use sparse matrices, and the computational operations were adapted to use sparse linear algebra routines.
We observed the expected reduction in memory requirements in our calculations, but our initial implementation was unable to substantially speed up the IT-IMSRG(2) solution over the IMSRG(2) solution.
Profiling and detailed benchmarks led us to suspect that this lack of performance is due to suboptimal handling of data around the sparse matrix format (e.g., random matrix element access in sparse matrices is slow, unlike with dense matrices).
These computational slow downs affected our implementation most heavily in the particle-hole part of the two-body commutator (see, e.g., Ref.~\cite{Herg16PR}), where the Pandya transformation is naturally implemented using unordered accesses in the input and output matrices.
However, given the speed ups observed in other parts of the solver when using sparse matrices (e.g., the particle-particle and hole-hole parts of the two-body commutator), we fully expect an optimized IT-IMSRG(2) solver to also reduce the computational cost of IMSRG(2) calculations in addition to the memory savings.

\section{Summary and Outlook}
\label{sec:outlook}

In this work, we introduced the importance-truncated IMSRG framework, where unimportant two-body matrix elements are truncated based on an IT measure and only the remaining elements are used for the many-body solution.
We investigated the systematics of this truncation, with the aim of identifying IT measures that allow for substantial compression ratios $R_C$ while introducing only small, controlled errors relative to the exact IMSRG(2) solution.

To this end, we employed the test case nucleus $^{40}$Ca and explored how truncating in different sub-blocks of the two-body Hamiltonian affects the IT error, how different IT measures perform, and how the IT truncation performs in model spaces of different sizes.
We found that restricting the IT truncation to only matrix elements with three or four particle indices (i.e., the hppp and pppp blocks) allows calculations up to reasonable compression ratios while introducing only small errors relative to the IMSRG(2) solution.
The different IT measures we explored all behaved systematically as a function of the compression ratio $R_C$, and we selected two measures, $\kappaX{\Gamma/\varepsilon}$ and $\kappaX{\Gamma n}$, to focus on based on their excellent performance when using soft Hamiltonians and relatively robust behavior when using harder Hamiltonians.
For both measures we observed that in larger model spaces the benefit of the IT-IMSRG in terms of compression substantially increased. Using soft Hamiltonians in the largest model spaces considered, we observed compression ratios of $R_C=100$ (i.e., 99\% of matrix elements are truncated) in calculations with less than 1~MeV error to the exact IMSRG(2) result.

Importance-truncated approaches typically attempt to approximately treat the truncated parts to correct for the truncation. In this work, we accounted for the truncated matrix elements via a third-order MBPT correction.
This correction serves as an estimate of the magnitude of the IT error, and for soft Hamiltonians its size is qualitatively consistent with the actual error observed.
For harder Hamiltonians, however, the error estimate systematically overestimates the actual error for intermediate and large compression ratios.
We understand this as a shortcoming of the perturbative treatment, which is expected to degrade when using harder Hamiltonians.

We extended our studies to medium-mass closed-shell nuclei ranging from $^{40}$Ca to $^{78}$Ni, comparing the harder, SRG-unevolved N$^3$LO 500 interaction to the soft, consistently SRG-evolved N$^3$LO 500 $\lambda=2.0$~fm$^{-1}$ interaction, and observed the same trends as in ${}^{40}\text{Ca}$. When using the SRG-evolved interaction, the IT-IMSRG(2) reproduces exact IMSRG(2) results excellently even at large compression ratios of $R_C=100$.
Moreover, the third-order MBPT corrections for the truncated matrix elements are consistent in magnitude to the actual observed errors, further supporting the use of this quantity as a reasonable IT error estimate for soft Hamiltonians.
When using the harder, SRG-unevolved Hamiltonian, larger errors to the exact IMSRG(2) results are observed, but the behavior is systematic and for smaller compression ratios the IT-IMSRG(2) results deviate from the IMSRG(2) results by only up to 1\%.

A first estimate of the computational benefits obtained by the IT approach in the IMSRG was studied by employing a sparse implementation of the IT-IMSRG solver. 
This adapted implementation used substantially less memory, profiting from the possible compression by the importance truncation,
but a direct improvement to the computational cost was not observed.
This suggests that a more fine-tuned IT-specific implementation is necessary to fully profit from the IT-IMSRG, and we expect that such an implementation would also see computational benefits at intermediate and large compression ratios.

All these results establish the IT-IMSRG as a promising tool to compress and accelerate many-body calculations.
These developments are of great interest for the ongoing efforts to develop the IMSRG(3)~\cite{Hein21IMSRG3},
where the computational costs of evaluating the most expensive commutators and the storage costs of handling three-body operators are strong limiting factors.
In preliminary studies on three-body matrix elements
using the generalized IT measure $\kappaX{\Gamma}$,
we have found that similar choices for $\kappaX{\Gamma}_{\text{min}}$ yield three-body compression ratios approximately an order of magnitude larger than the two-body compression ratios obtained in this work.
Substantial reduction ($R_C \sim 40$-$200$) in the number of three-body matrix elements could bring memory requirements associated with three-body operators into the range accessible with standard supercomputing nodes.
An IT-IMSRG(3) solver tuned to handle the extremely sparse structure of the resulting operators could extend the range of the many-body method and make large model-space truncations or more expensive approximate IMSRG(3) truncations accessible.

We expect that the IT-IMSRG approach can also be adapted to work with extensions of the IMSRG that target open-shell systems, such as the multireference or the valence-space IMSRG, with appropriately selected importance measures.
Some of the proposed measures in this work rely on low-order pertubative estimates and are thus not directly applicable in open-shell systems.
On the other hand, occupation-based IT measures (with occupation numbers obtained from an appropriate one-body density calculation) might be straightforwardly applied in a multireference framework, and the derivative-based measure is directly generalizable to other IMSRG frameworks.
Open-shell-specific IT measures could be constructed by taking information from symmetry-breaking and projection calculations or considering details of the valence space.

Extending the IT approach to other operators and observables in the IMSRG is a next natural step, where either the same measure as for the Hamiltonian can be applied or new operator-specific measures can be developed.

\acknowledgments

We thank P.~Arthuis and L.~Zurek for useful discussions and C. Johnson for help in setting up the BIGSTICK code.
This work was supported in part by the  Deutsche  Forschungsgemeinschaft  (DFG,  German Research Foundation) -- Project-ID 279384907 -- SFB 1245 and by the European Research Council (ERC) under the European Union’s Horizon 2020 research and innovation programme (Grant Agreement No.~101020842).

\bibliography{strongint}

\end{document}